\documentclass[aps,prl,reprint]{revtex4-1}
\usepackage[usenames,dvipsnames]{color}
\usepackage{amsmath,array,amssymb}
\usepackage{graphicx}
\usepackage{hyperref}
\usepackage{textcomp}
\usepackage{braket}
\usepackage{multirow}
\usepackage{indentfirst}
\usepackage[FIGTOPCAP,raggedright,nooneline,bf,footnotesize]{subfigure}
\usepackage{paralist}
\usepackage{overpic}
\usepackage{soul} 
\usepackage{xspace} 

\newcommand{\cmt}[1]{}

\renewcommand{\vec}[1]{\boldsymbol{#1}}

\renewcommand{\eqref}[1]{Eq.~(\ref{#1})}

\newcommand{\eryso}[0]{{Er$^{3+}$:Y$_2$SiO$_5$}\xspace}
\newcommand{\yso}[0]{Y$_2$SiO$_5$\xspace}
\newcommand{\er}{Er$^{3+}$\xspace}
\newcommand{\y}{Y$^{3+}$\xspace}

\graphicspath{{img/}}

\usepackage{pdfpages}

\makeatletter
\AtBeginDocument{\let\LS@rot\@undefined}
\makeatother

\begin{document}

\title{Selective optical addressing of nuclear spins through superhyperfine interaction in rare-earth doped solids}

\author{B. Car}
\author{L. Veissier}
\author{A. Louchet-Chauvet}
\author{J.-L. Le Gou\"et}
\author{T. Chaneli\`ere}
\affiliation{Laboratoire Aim\'e Cotton, CNRS, Universit\'e Paris-Sud, ENS Paris-Saclay, Universit\'e Paris-Saclay,  91405 Orsay, France}

\date{\today}

\begin{abstract}

In \eryso, we demonstrate the selective optical addressing of the $^{89}$Y$^{3+}$ nuclear spins through their superhyperfine coupling with the \er electronic spins possessing large Land\'e $g$-factors. We experimentally probe the electron-nuclear spin mixing with photon echo techniques and validate our model. The site-selective optical addressing of the \y nuclear spins is designed by adjusting the magnetic field strength and orientation. This constitutes an important step towards the realization of long-lived solid-state qubits optically addressed by telecom photons.

\end{abstract}


\maketitle

Nuclear spins in solids represent excellent systems for storing and processing quantum information because of their long coherence lifetimes, coming from their limited exposure to environmental fluctuations. During the last decade, impressive progress has been made towards the coherent manipulation of nuclear spins and the control of their interaction with the environment, which is crucial to achieve long-lived solid-state qubits \cite{childress2006coherent}. For this purpose, the superhyperfine interaction between an electron spin and a neighboring ligand nuclear spin has known a renewed interest since it offers an efficient way of accessing nuclear spins. Indeed, the two spins live in symbiosis. The electron spin can be strongly excited by RF or optical fields to produce well-defined quantum states in a variety of different solid-state systems as NV \cite{maurer_room-temperature_2012, yang2016high} or SiV \cite{SiV1,SiV2} centers in diamond, quantum dots in semiconductors \cite{QD} and donors in silicon as phosphor \cite{steger_quantum_2012}, bismuth \cite{morley2013quantum} or the optically active selenium \cite{Morsee1700930}. The information can then be mapped into the ligand nuclear spin to realize long-lived qubits memories with lifetimes up to minutes in impurity-doped solids \cite{maurer_room-temperature_2012, steger_quantum_2012}. Material purification in order to avoid any other nuclear spins in the medium, which would lead to decoherence of the stored qubits, is though often required.

In this context, \eryso is particularly interesting as it offers an environment with minimized magnetic moments : $^{89}$Y ($I=1/2$) is the only nuclear spin with a single stable isotope, making the surrounding nuclei quite equivalent. Moreover, the electronic spin of \er possesses a large Land\'e $g$-factors, enabling strong interaction with RF excitation, even in the regime where superconducting qubits operate \cite{probst, bienfait2016reaching}. \er -doped crystals are also known because the $^{4} I_{15/2} \rightarrow \, ^{4}I_{13/2}$ optical transition falls in the telecom range, with homogeneous linewidths narrower than 100~Hz \cite{bottger_effects_2009}. Additionally, optical addressing of single Er ions has been recently achieved in this material via the coupling with silicon nanophotonic structures \cite{dibos_isolating_2017}. The complexity comes from the low symmetry of the \yso crystalline structure. Previous studies derived from electron paramagnetic resonance (EPR) techniques \cite{Mims_SHF1, Mims_SHF2,mims1965pulsed} have indeed revealed a profusion of inequivalent neighbor nuclear spins interacting with the erbium spins \cite{guillot-noel_direct_2007, Ahlefeldt20101594}, which is a major drawback to control the electron-nuclear coupling.

In this work, we demonstrate that a single class of yttrium nuclear spins can be optically addressed through the erbium telecom transition, despite a large number of surrounding yttrium ions. We theoretically calculate the superhyperfine interaction from the relative positions of the yttrium ions. A full mixing between electron and nuclear states appears for specific orientations and strengths of the external magnetic field, and at specific locations in the crystalline cell. Because of this mixing, the nuclear spin states form an optical $\Lambda$-system, an actively pursued feature in qubit design to perform optical pumping or spin state initialization. We give a comprehensive analytical study of the electron-nuclear mixing in the specific case of a low site symmetry where the Zeeman $g$-tensor is highly anisotropic, and then experimentally probe the superhyperfine interaction with photon echo techniques. A strong modulation due to the Er-Y coupling is analyzed and successfully compared with the theoretical model. This work can be directly transposed to other rare-earth Kramers ion doped crystals, some of them being actively investigated for quantum information \cite{tiranov_temporal_2016,zhong_interfacing_2017}.

We consider the interaction of the erbium ion electron spin with the most abundant surrounding nuclear spins, namely $^{89}$Y ($I=1/2$). In a \eryso crystal, one naturally finds only 4.7\% of $^{29}$Si, 0.04\% of $^{17}$O. The optical addressing of the nuclear spins will be mediated by the optical excitation of erbium ions on the $^{4}I_{15/2} \rightarrow ^{4}I_{13/2}$ zero-phonon line. We specifically use a low doping concentration (10 ppm) crystal to avoid the so-called erbium spin flip-flops in the regime of small external magnetic fields \cite{bottger_optical_2006}. As a consequence, the spectral diffusion is significantly reduced, allowing us to observe optical coherence lifetimes up to 200~$\mu$s even at magnetic fields below 100~mT. This range is particularly interesting because the superhyperfine interaction is here comparable to the nuclear Zeeman splitting, precisely leading to a strong electron-nuclear mixing, as described in the following. The response of the $^{167}$Er isotope (22\% of the dopant concentration, with a nuclear spin of $7/2$) is broadly spread over a large amount of possible hyperfine transitions \cite{guillot-noel_hyperfine_2006} and therefore can be neglected because of the optical selection. 

\begin{figure}[t]
\centering
\includegraphics[width=0.9\columnwidth]{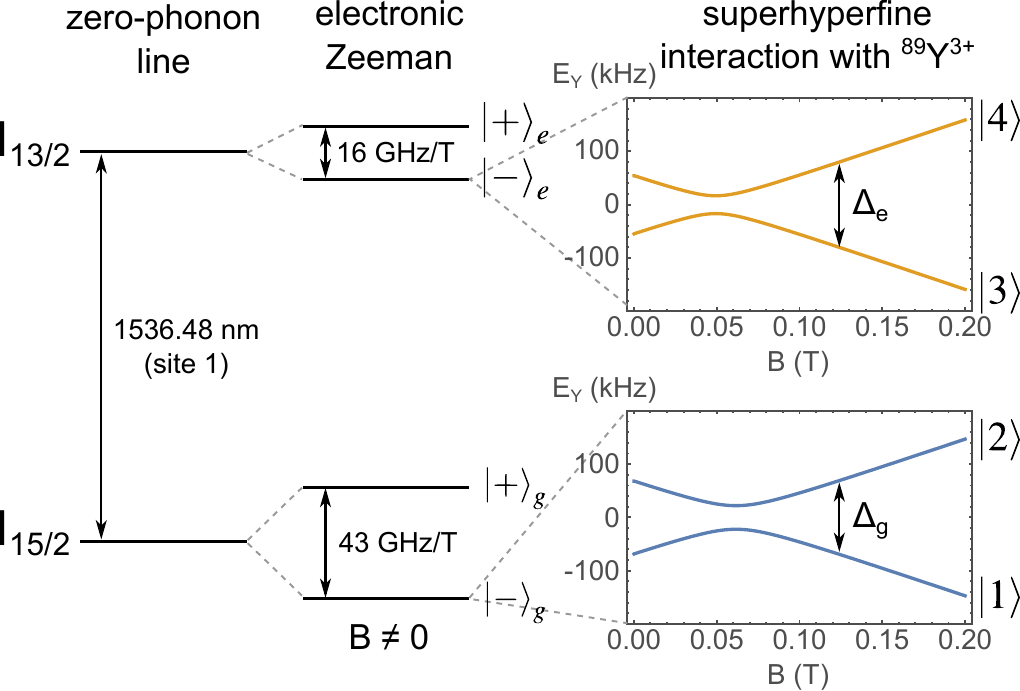}
\caption{Relevant energy structure of the \er ion in the \yso matrix. The application of a magnetic field lifts the degeneracy of the doublets in the ground and excited states of the zero-phonon line via electronic Zeeman interaction, leading to Zeeman coefficients of 43 and 16~GHz/T in respectively the ground and excited states for a magnetic field oriented in the $D_1$-$D_2$ plane at 225$^\circ$ from $D_1$ (see main text). The superhyperfine coupling between \er and nuclear \y spins splits each Er$^{3+}$ state into a nuclear doublet at low field. The linear behavior of the nuclear Zeeman interaction leads to avoided crossings in the \y energy spectra $E_{\rm Y} \left( B \right)$, as shown in the insets for a specific \y ion for which $r=5.46$~{\AA}.}
\label{fig:system}
\end{figure}

For a given state of the \er ion, labeled  $g$ or $e$ for respectively $^{4}I_{15/2}$ or $^{4}I_{13/2}$, the total $4 \times 4$ Hamiltonian for the erbium spin coupled to a single yttrium nuclear spin can be written as
\begin{equation}
H^{\rm tot}_{g,e} = - \vec{\mu}^{\rm Er}_{g,e} \cdot \vec{B} - \vec{\mu}_{\rm Y} \cdot \vec{B} + H_{g,e}^{\rm Er-Y} \;,
\label{eq:Heff}
\end{equation} 
where $\vec{\mu}^{\rm Er}_{g,e} $ is the \er electronic spin in the ground or excited state, $\vec{\mu}_{\rm Y}$ is the \y nuclear spin, $\vec{B}$ the externally applied magnetic field, and $H_{g,e}^{\rm Er-Y}$ the magnetic dipole-dipole electron-nuclear interaction. The first term of Eq.~\ref{eq:Heff}, the electronic Zeeman coupling of \er spins, splits the ground and excited state doublets by several GHz for $B=100$~mT, with eigenstates $\left\lbrace \ket{+},\ket{-} \right\rbrace_{g,e}$ as shown in Fig.~\ref{fig:system}. Indeed, the gyromagnetic ratios of the \er spins in the ground and excited states are exceptionally large, ranging from 15 to 150~GHz/T depending on the magnetic field orientation, which is 4 and 5 orders of magnitude larger than the yttrium nuclear spin (2.1~MHz/T). Thus, we treat the last two terms as perturbation and replace the erbium magnetic moment in $H_{g,e}^{\rm Er-Y}$ by its expectation value $\langle \vec{\mu}^{\rm Er}_{g,e} \rangle$ on the $\left\lbrace \ket{+},\ket{-} \right\rbrace_{g,e}$ eigenstates. In consequence, the perturbation $2 \times 2$ Hamiltonian $H_{g,e}^\prime$ for the \y spin is given by
\begin{align}
H_{g,e}^\prime  = - \vec{\mu}_{\rm Y}  \cdot \left( \vec{B} - \frac{\mu_0}{4\pi} \left[ \frac{\langle \vec{\mu}^{\rm Er}_{g,e} \rangle }{r^3}-3\frac{\left(\langle \vec{\mu}^{\rm Er}_{g,e} \rangle  \cdot\vec{r}\right)\cdot\vec{r}}{r^5}\right]  \right) \; ,
\label{Hperturb}
\end{align}
where $\mu_0$ is the vacuum permeability and $\vec{r}$ the vector joining the two spins ($r$ is the distance).

The relevant parameter characterizing the optical excitation of the \y nuclear spin is the branching ratio $R=\frac{|\braket{2|3}|^2}{|\braket{1|3}|^2} = \frac{|\braket{1|4}|^2}{|\braket{2|4}|^2}$ between the eigenstates of $H_{g,e}^\prime$, namely the superhyperfine levels $\ket{1}$ and $\ket{2}$ (resp. $\ket{3}$ and $\ket{4}$) in the ground state $\ket{-}_g$ (resp. excited state $\ket{-}_e$), as defined in Fig.~\ref{fig:system}. We additionally introduce the branching contrast $\rho$ directly connected to $R$ as
\begin{equation}
\rho=\frac{4R}{(1+R)^2} \; ,
\label{eq:rho_def}
\end{equation}
which characterizes the degree of spin mixing that is possible to achieve optically. When $\rho=1$, the two optical branches of the $\Lambda$-system are equally probable: the \y spin can be addressed optically and prepared in an equally weighted superposition state ($ \frac{ \ket{1} +\ket{2} } {\sqrt{2}} $ for instance). As soon as the two transitions probabilities differ, $\rho$ decreases.

The perturbative expansion is actually much more than a formal simplification. Because the erbium spin energy dominates the superhypefine and nuclear Zeeman interaction, the nuclear spin mixing is solely explained by the change of the expectation values $\langle \vec{\mu}^{\rm Er}_{g,e} \rangle$ from the ground to the excited state of erbium. $\langle \vec{\mu}^{\rm Er}_{g} \rangle$ and $\langle \vec{\mu}^{\rm Er}_{e} \rangle$ are never exactly aligned because of the Er$^{3+}$ strongly anisotropic $g$-tensors \cite{sun_magnetic_2008}. The perturbation Hamiltonian can indeed be alternatively rewritten from \eqref{Hperturb} as $H_{g,e}^\prime  = - \vec{\mu}_{\rm Y} \cdot \vec{B}_{g,e} \left( \vec{r} \right)$ where $\vec{B}_{g,e}$ is the total magnetic field 
seen by the \y spin (location $\vec{r}$), including the magnetic field 
generated by the \er spin of moment $\langle \vec{\mu}^{\rm Er}_{g,e} \rangle$. Following this interpretation, the electron-nuclear mixing appears when the total field is strongly modified by the optical excitation of the \er ion. More precisely, it is maximized when $\vec{B}_{g} \left( \vec{r} \right) \perp \vec{B}_{e} \left( \vec{r} \right)$. Indeed, the branching ratio is given by 
\begin{equation}
R = \tan^2 \left( \alpha /2 \right) \; ,
\label{eq:branchingratio}
\end{equation}
where $\alpha$ is the angle between $ \vec{B}_{g}$ and $\vec{B}_{e}$, which gives $\rho = \sin^2 \left( \alpha \right)$ \cite{SM_rho}.
The appearance of avoided crossings on the \y spin spectra in Fig.~\ref{fig:system} occurring at different magnetic field strengths for the ground and excited state of erbium is actually the blueprint of the optically induced vectorial tilt of $\vec{B}_{g}$ and $\vec{B}_{e}$. \eqref{eq:branchingratio} also reminds us that the branching ratio does not depend on the gyromagnetic factor of the nuclear spin but only on its position $\vec{r}$.

Because a large branching contrast $\rho$ requires a maximum variation of the total magnetic field, it only appears at certain specific locations $\vec{r}$ in the crystal cell. This is the key idea leading to the optical selectivity of the nuclear spin addressing. For a given magnetic field orientation, the magnetic moment of erbium is fixed. The branching contrast is then given by the magnetic fields $\vec{B}_{g,e}$ following \eqref{eq:branchingratio} or equivalently by diagonalizing the pertubative Hamiltonians $H_{g,e}^\prime$ of \eqref{Hperturb}.

\begin{figure}[t]
\includegraphics[width=0.9\columnwidth]{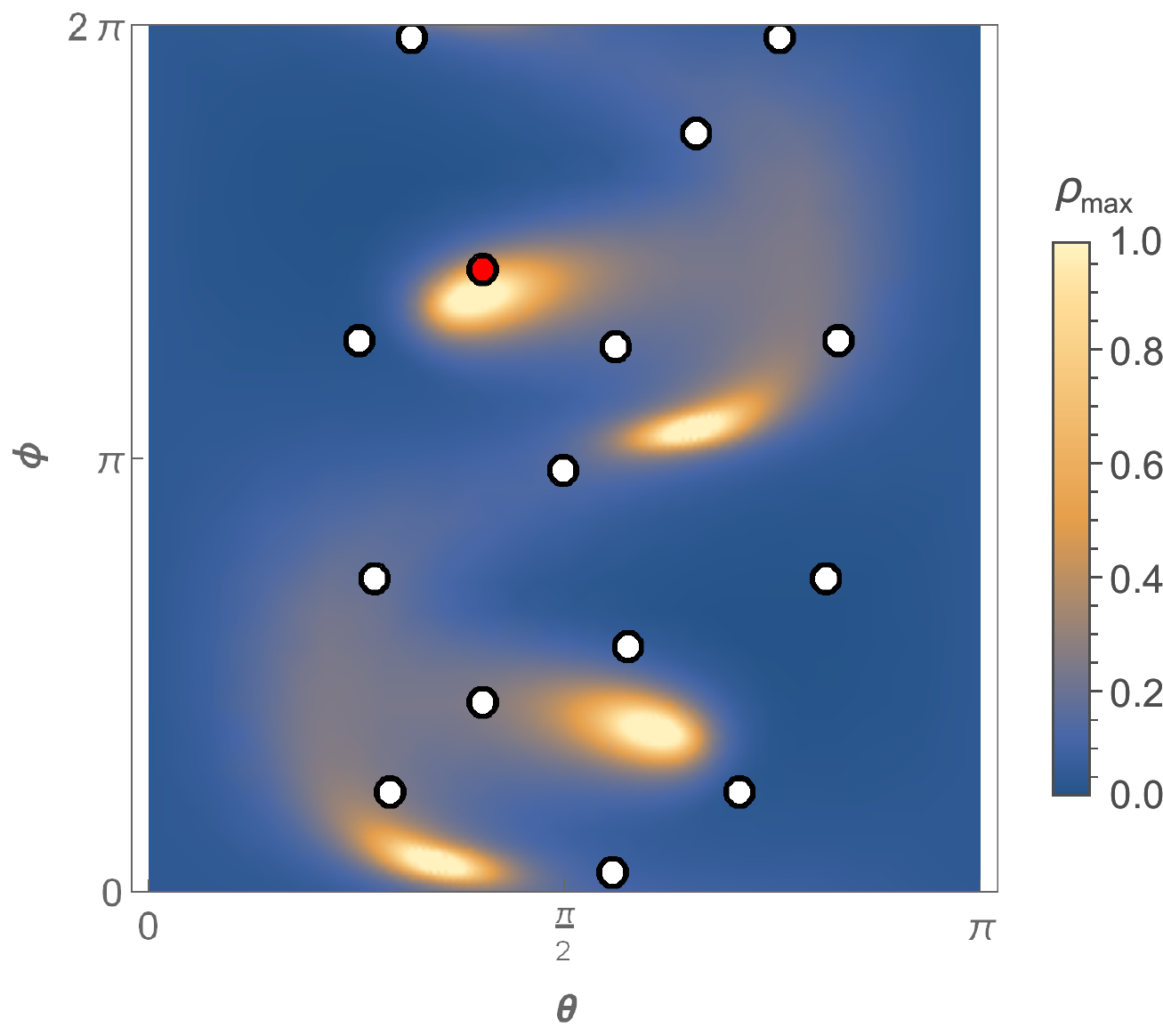}
\centering
\caption{Map of the maximum branching contrast $\rho_{\rm max}$ as a function of the polar angle $\theta$ and the azimuthal angle $\phi$ of $\vec{r}$ in the $(D_1,D_2,b)$ crystal frame \cite{YSOstructure}. We consider the external magnetic field at $225^{\circ}$ from $D_1$, and \er ions at site 1 and orientation A \cite{SM1}. The values of the $g$-tensors can be found in \cite{sun_magnetic_2008}. The dots represent the positions of the fifteen nearest \y ions from the \er ion (distances from 3.40~{\AA} to 5.74~{\AA}). The \y ion with cartesian coordinates (-1.01,-5.11,1.64)~{\AA} is pinned in red. }
\label{fig:map_rho}
\end{figure}

We identify a particularly interesting configuration in which \y nuclear spins can be strongly coupled to \er ions.
This occurs when the magnetic field $\vec{B}$ is oriented at $225^{\circ}$ from $D_1$ within the $(D_1,D_2)$ plane, $D_1$ and $D_2$ being the optical extinction axes of \yso. 
For given angular coordinates of the \y position $\vec{r}$, the branching contrast $\rho$ reaches a maximum $\rho_{\rm max}$ as a function of the magnetic field strength. Fig.~\ref{fig:map_rho} shows the spatial mapping (angular coordinates) of $\rho_{\rm max}$ for this specific magnetic field orientation, and for \er ions of orientation (magnetic sub-site) A in site 1 \cite{SM1}. The $\rho_{\rm max}$ map is composed of well isolated peaks, highlighting the strong selectivity of the \y ions optical addressing. The value of $\rho_{\rm max}$ is independent of the \y ion distance $r$ from the \er center but the field strength maximizing $\rho$ crucially depends on $r$. By slightly varying the orientation or the strength of the field close to the maximum value of the branching contrast, the optical addressing of the \y spins can be activated or inhibited.

After positioning the nearest \y ions \cite{Maksimov1970} on the $\rho_{\rm max}$ map, one can notice that only one \y ion ($r=5.46$~{\AA}) is positioned close to a maximum of $\rho_{\rm max}$ (in red in Fig.~\ref{fig:map_rho}), meaning that its flipping probability can be maximized for an appropriate magnetic field strength (oriented at $225^{\circ}$ from $D_1$). In the following, we experimentally study the specific interaction of \er ions with this \y ion using photon echo techniques in order to measure the corresponding energy splittings and branching contrast.

For this purpose we cool down to $1.8$~K a 10 ppm \eryso crystal grown by Scientific Materials Corporation. 
The light propagates along the $b$-axis of the crystal \cite{SM2}. We perform 2-pulse photon echo measurements on the lowest to lowest spin state ($\ket{-}_g \leftrightarrow \ket{-}_e$)  transition of site 1 (1536.38~nm) to optically observe the superhyperfine interaction with a kHz resolution. This precision is much narrower than the typical inhomogeneous broadening ($\sim$ 500~~MHz). The short pulse excitation bandwidth should cover the superhyperfine splittings \cite{SM2}.

\begin{figure}[t]
\centering
\includegraphics[width=0.9\columnwidth]{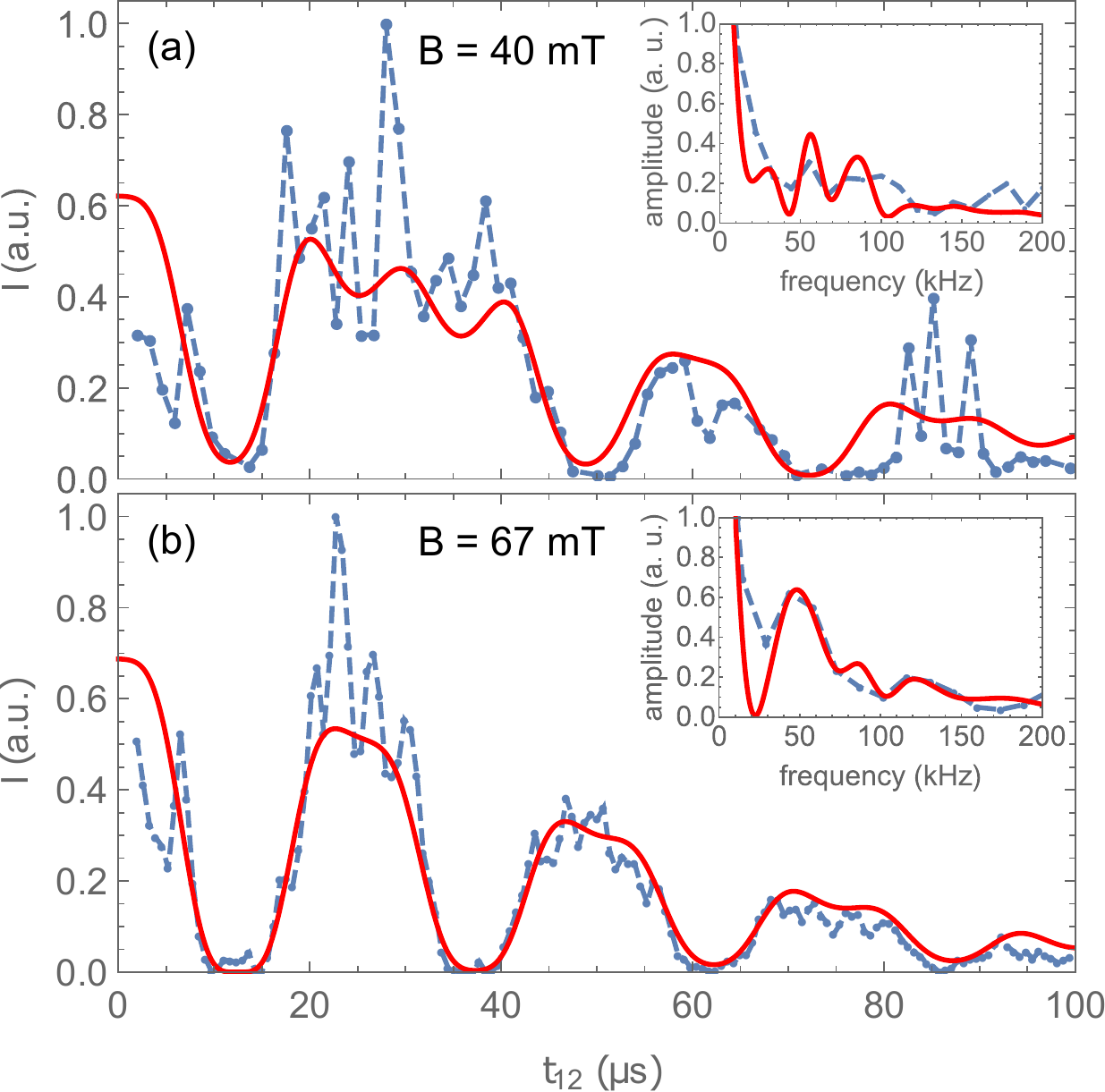}
\caption{Modulated 2-pulse photon echo on the $\ket{-}_g \leftrightarrow \ket{-}_e$ transition of \er ions of site 1. The temperature is $1.8$~K and the magnetic field is oriented at $225^{\circ}$ from $D_1$ with strength of (a) 40~mT and (b) 67~mT. The experimental data are fitted using Eq.~\ref{eq:modulation} (red lines) to extract the superhyperfine transition frequencies $\Delta_{g,e}$ and the branching contrast $\rho$. The insets show the corresponding spectra, i.e. Fourier transforms of our experimental echo decays (blue dots) and of the models (red lines).}
\label{fig:2PE}
\end{figure}

By varying the delay $t_{12}$ between the excitation pulses, we observe strong modulations in the emitted echo intensity revealing the strong spin mixing, as shown by Fig.~\ref{fig:2PE}. Following Mitsunaga's theory \cite{mitsunaga_cw_1990} restricted to a single nuclear spin coupling, the echo intensity can be written as
\begin{align}
\begin{split}
I \left(t_{12}\right) = & I_0 \, \exp \left[ -2 \left(\frac{2 \, t_{12}}{T_2} \right)^x \right] \\
&\times \left\lbrace 1-\frac{\rho}{2} \left[ 1-\cos\left(2 \pi \Delta_g t_{12}\right) \right] \left[ 1-\cos \left(2 \pi \Delta_e t_{12} \right) \right] \right\rbrace ^2
\label{eq:modulation}
\end{split}
\end{align}
where $T_2$ is the optical coherence lifetime and $x$ the Mims exponent, which accounts for spectral diffusion processes \cite{bottger_optical_2006}. The parameters $\Delta_g$ and $\Delta_e$ are respectively the superhyperfine splittings of the states $\ket{-}_g$ and $\ket{-}_e$ respectively (see Fig.~\ref{fig:system}).

Fig.~\ref{fig:2PE} shows the experimental echo decays for $B=40$ and 67~mT. The strong echo modulations are well reproduced by Eq.~\ref{eq:modulation} allowing to extract $\rho$ as a fitting parameter. The Mims exponent is fixed to $x=1.5$, according to previous studies \cite{bottger_optical_2006}. For $B=40$~mT, we find $\Delta_g=49$~kHz and $\Delta_e=33$~kHz. At $B=67$~mT, both splittings become almost equal to $\Delta_g=\Delta_e=41$ kHz. We also observe an underlying fast modulation, which is attributed to the interaction with an \y ion located at a distance of 3.72~{\AA}, and for which we calculate $\Delta_g = 260$ kHz and $\Delta_e = 231$ kHz, with $\rho = $ 0.19. In order to visualize our spectral resolution, we calculate the Fourier transforms of our experimental data, as well as the fitting models, as shown in the insets of Fig.~\ref{fig:2PE}. We identify the presence of one main peak for $B=67$~mT when both splittings are almost equal, and two peaks at $B=40$~mT.

\begin{figure}[t]
\centering
\includegraphics[width=1\columnwidth]{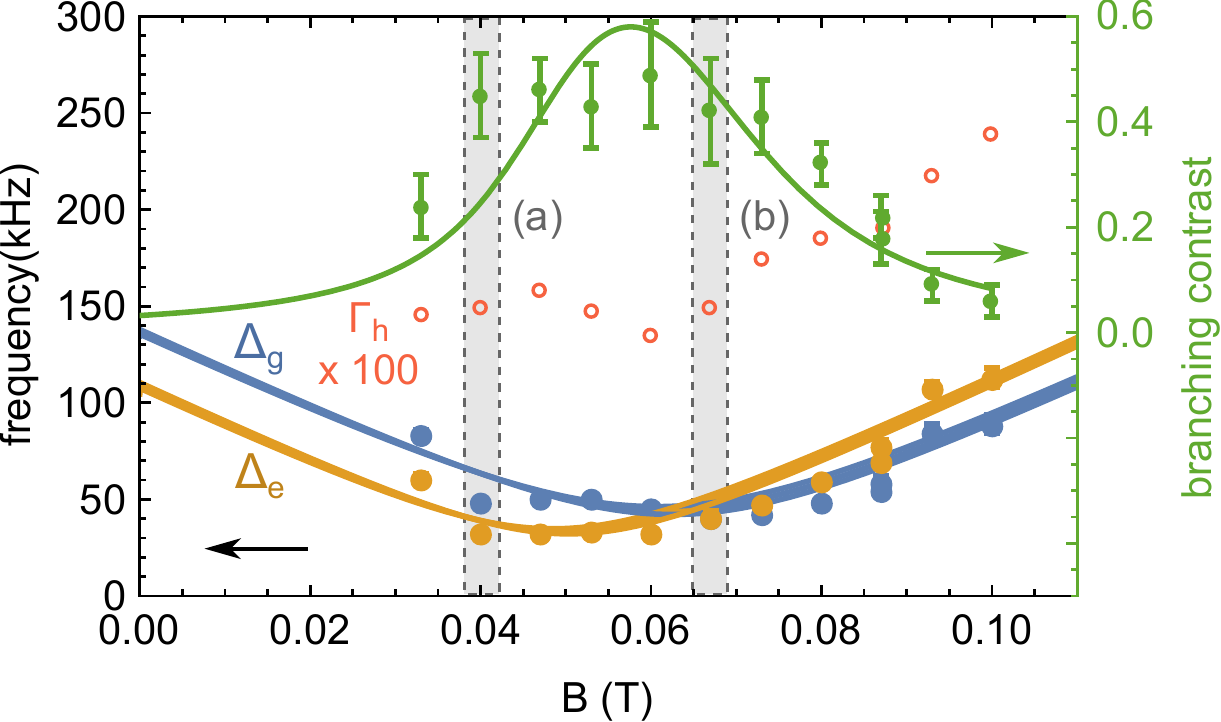}
\caption{Superhyperfine transition frequencies $\Delta_{g,e}$ and branching contrast $\rho$ as a function of the magnetic field strength $B$ for $\vec{B}$ oriented at $225^{\circ}$ from $D_1$. Experimental points and their error bars were extracted from the fit of echo decays (Eq.~\ref{eq:modulation}) and the shaded areas labeled (a) and (b) highlight the values corresponding to Fig.~\ref{fig:2PE} ($B=40$ and 67~mT). The solid lines are the calculated values. The homogeneous linewidth $\Gamma_{\rm h} \times 100$ of the \er ion optical transition is represented by the red empty circles.}
\label{fig:modfreq_Bdep}
\end{figure}

We experimentally follow the variations of $\Delta_g$, $\Delta_e$ and $\rho$ as a function of the external magnetic field strength $B$. The comparison with the theoretical predictions is shown in Fig.~\ref{fig:modfreq_Bdep}. The width of the solid lines (theoretical calculation) accounts for the uncertainties in the magnetic field orientation and strength, as well as for the estimated inhomogeneities along the 8~mm long crystal. The good agreement validates our model of the superhyperfine interaction that leads to a strong electron-nuclear mixing. Moreover, as expected the branching contrast drastically varies with the magnetic field strength. Thus, we demonstrate here the possibility to effectively tune the Er-Y coupling by changing the external magnetic field by only a few tens of mT, which is experimentally easily achievable.

The homogeneous optical linewidth $\Gamma_{\rm h} = 1/\left( \pi T_2 \right)$, also extracted from the fit of Eq.~\ref{eq:modulation} and shown on Fig.~\ref{fig:modfreq_Bdep}, is a key parameter for the observation of the superhyperfine coupling. Benefiting from the low concentration of our sample, we measure linewidths between 1.4~kHz and 2.4~kHz, so always much smaller than the observed splittings. This allows the optical selective excitation of the \y nuclear spins. 

Our case may be perceived as particular because of the exceptionally large $g$-factor of erbium, but together with the small nuclear moment of \y, the induced energy shifts are typical of the superhyperfine interaction in solids such as chromium electron spin coupled to aluminum in Ruby \cite{PhysRevB.7.1834_ruby}, N-V center coupled to a nearby $^{13}$C in diamond \cite{van1990optically} or rare-earth paramagnetic impurity coupled to non-Kramers rare-earth ligands \cite{Ahlefeldt20101594}. Therefore, our observations also point out the interest of optical measurements to investigate the electron-nuclear coupling, usually interrogated via Electron Spin Echo Envelope Modulation (ESEEM) \cite{Mims_SHF1, Mims_SHF2,guillot-noel_direct_2007} and Pulsed electron nuclear double resonance (ENDOR) techniques \cite{mims1965pulsed} derived from EPR. Optics totally relaxes the constraints on the magnetic field values and allows to precisely investigate the region where the mixing between electron and nuclear spin states is maximum, usually well below the EPR X-band. Not relying on population difference in the spin states, the present all-optical method can be implemented more easily than optically-detected magnetic resonance (ODMR) techniques in the case of structures smaller than the optical inhomogeneous broadening.

To conclude, we achieve an optically selective excitation of nuclear spins in the \yso matrix via superhyperfine interaction with \er ions. A good understanding and control of the superhyperfine interaction is demonstrated. The electron-nuclear mixing is revealed by strong modulations in photon echo measurements, with modulation frequencies and amplitude matching our theoretical calculations. Despite the complexity of the low symmetry \yso crystal, we accurately model the interaction between an electron spin qubit and neighboring nuclei. Moreover, the understanding of the interaction between impurities and ligands, an underlying physical ingredient of the futur solid-state quantum devices, will support the identification of the sources of decoherence for both the electron and nuclear spins \cite{PhysRevB.82.201201, PhysRevB.78.094303}. We anticipate the nuclear spin coherence lifetime to be directly limited by the Er spin relaxation, for which values as long as 4 s has been observed \cite{probst_anisotropic_2013}. Several techniques can be implemented in order to reduce spin decoherence, as suggested for silicon or diamond nuclear spin bath: dynamic nuclear polarization \cite{PhysRevLett.114.247603, scheuer2016optically} or advanced material development \cite{tyryshkin2012electron}.
Finally, this work paves the way for the optical control of long-lived solid-state qubits in \eryso. The interplay between the optical and the spin properties in erbium doped materials also opens the perspective of a unit quantum efficiency modulator to coherently up-convert microwave photons to the optical telecom domain \cite{obrien_interfacing_2014, fernandez-gonzalvo_coherent_2015, PhysRevLett.113.203601, kukharchyk2017optical}. This latter appears as a candidate to link local quantum processing nodes \cite{ladd_quantum_2010} and quantum communication channels \cite{hensen2015loophole}.

We received funding from the national grant ANR DISCRYS (ANR-14-CE26-0037-02), from Investissements d'Avenir du LabEx PALM ExciMol and OptoRF-Er (ANR-10-LABX-0039-PALM). The research leading to these results has received funding from the People Programme (Marie Curie Actions) of the European Union’s Seventh Framework Programme (FP7/2007-2013) under REA grant agreement n. PCOFUND-GA-2013-609102, through the PRESTIGE
programme coordinated by Campus France.

This work is dedicated to the memory of our colleague Daniel Ricard.

\pagebreak
\widetext
\begin{center}
\textbf{\large Supplemental Materials: Selective optical addressing of nuclear spins through superhyperfine interaction in rare-earth doped solids}
\end{center}
\setcounter{equation}{0}
\setcounter{figure}{0}
\setcounter{table}{0}
\makeatletter
\renewcommand{\theequation}{S\arabic{equation}}
\renewcommand{\thefigure}{S\arabic{figure}}
\renewcommand{\bibnumfmt}[1]{[S#1]}
\renewcommand{\citenumfont}[1]{S#1}

\section{Derivation of the branching ratio}

We derive here the expression of $R$ and $\rho$, respectively the branching ratio and contrast for the optical addressing of \y spins, as a function of the angle $\alpha = \left( \vec{B}_{g},\vec{B}_{e} \right)$ between the two total magnetic fields, one in the case of the \er spin in the ground state ($\vec{B}_{g}$), and the other in the case of the \er spin in the excited state ($\vec{B}_{e}$). When the \er spin is in $\ket{-}_g$, the eigenstates $\ket{1}$ and $\ket{2}$ of the perturbative Hamiltonian $H'_g$ are calculated in the orthonormal frame $\left( \hat{i},\hat{j},\hat{k} \right)$ with $\hat{k} \parallel \vec{B}_{g}$ as quantization axis. Similarly, when the \er spin is in $\ket{-}_e$, the eigenstates $\ket{3}$ and $\ket{4}$ of $H'_e$ are calculated in the orthonormal frame $\left( \hat{i}',\hat{j}',\hat{k}' \right)$ with $\hat{k}' \parallel \vec{B}_{e}$. Then, to change basis, we use the Wigner $D$-matrix, which, for a $1/2$ spin, is given by
\begin{equation}
D^{1/2} \left( \alpha, \beta, \gamma \right) = \begin{pmatrix}
e^{-  i \alpha/2} \cos \left( \frac{\beta}{2} \right) e^{- i \gamma /2} & - e^{- i \alpha/2} \sin \left( \frac{\beta}{2} \right) e^{ i \gamma/2} \\
e^{i \alpha/2} \sin \left( \frac{\beta}{2} \right) e^{- i \gamma/2} & e^{ i \alpha/2} \cos \left( \frac{\beta}{2} \right) e^{  i \gamma/2}
\end{pmatrix} \; ,
\label{eq:rotmat}
\end{equation}
where $\alpha$, $\beta$, $\gamma$ are the Euler angles of the rotation. Thus, we can write the eigenstates $\ket{3}$ and $\ket{4}$ as function of $\ket{1}$ and $\ket{2}$, as follows
\begin{align}
\ket{3} = & e^{-  i \alpha/2} \cos \left( \frac{\beta}{2} \right) e^{- i \gamma /2} \ket{1} - e^{- i \alpha/2} \sin \left( \frac{\beta}{2} \right) e^{ i \gamma/2} \ket{2} \; , \\
\ket{4} = & e^{i \alpha/2} \sin \left( \frac{\beta}{2} \right) e^{- i \gamma/2} \ket{1} + e^{ i \alpha/2} \cos \left( \frac{\beta}{2} \right) e^{  i \gamma/2} \ket{2} \; .
\end{align}
Then, one can see easily that
\begin{equation}
R=\frac{|\braket{2|3}|^2}{|\braket{1|3}|^2} = \frac{|\braket{1|4}|^2}{|\braket{2|4}|^2} = \tan^2 \left( \beta / 2 \right) \; .
\end{equation}
The Euler angle $\beta$ is actually the angle between $\hat{k}$ and $\hat{k'}$, so $\beta=\alpha$, and $R=\tan^2 \left( \alpha / 2 \right)$. Finally, for the branching contrast, we obtain $\rho =\sin^2 \left( \alpha \right)$.

\section{Simulations for \er ions of site 1, orientation B}

\begin{figure}[t]
\includegraphics[width=0.5\columnwidth]{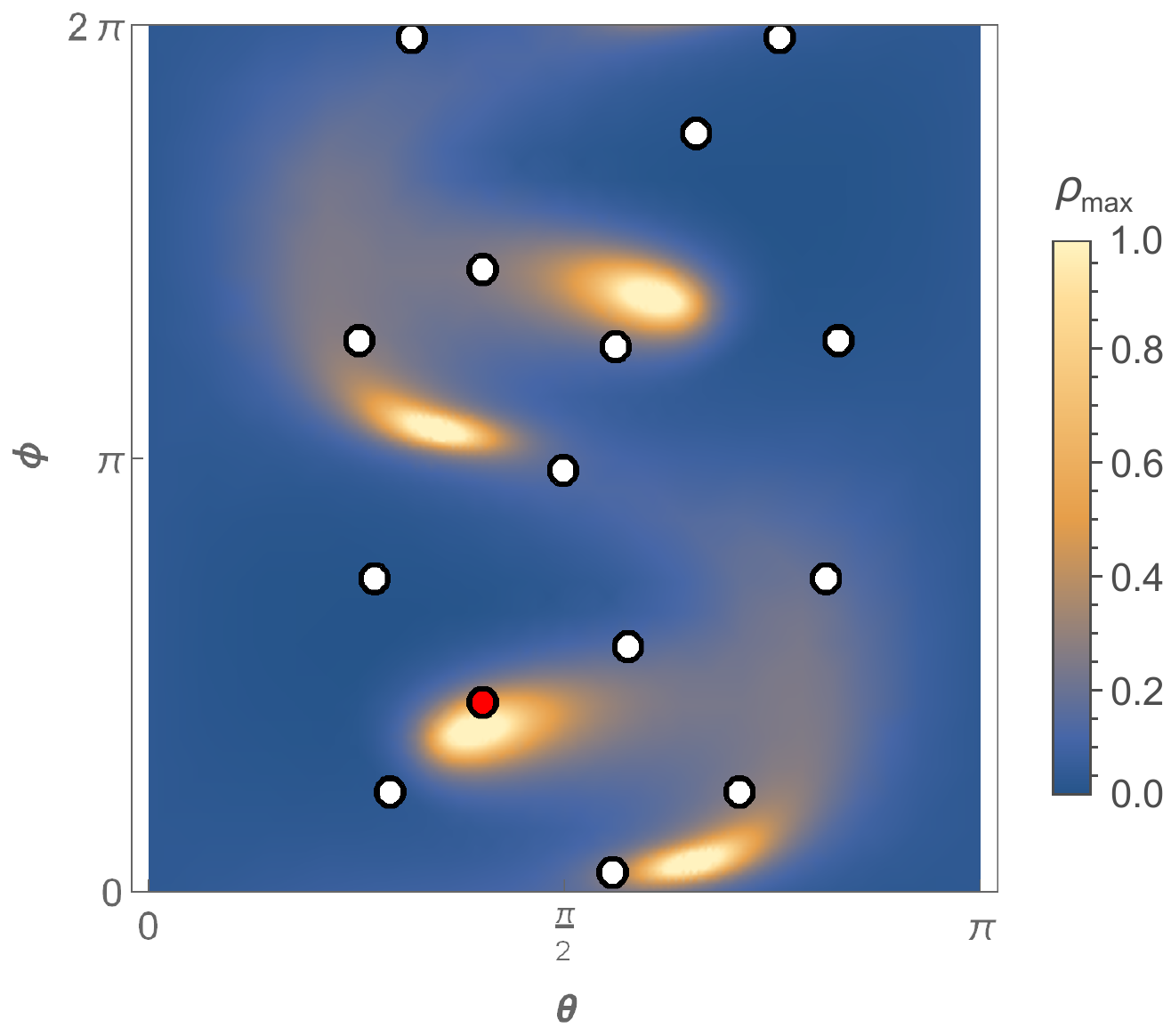}
\centering
\caption{Map of the maximum branching contrast $\rho_{\rm max}$ as a function of the polar angle $\theta$ and the azimuthal angle $\phi$ of $\vec{r}$ in the $(D_1,D_2,b)$ crystal frame \cite{YSOstructure_SM}. We consider the external magnetic field at $225^{\circ}$ from $D_1$, and \er ions at site 1 and orientation B (see main text for definition). The values of the $g$-tensors can be found in \cite{sun_magnetic_2008_SM}. The dots represent the positions of the fifteen nearest \y ions from the \er ion (distances from 3.40~{\AA} to 5.74~{\AA}). The \y ion with cartesian coordinates (1.01,5.11,1.64)~{\AA} is pinned in red.}
\label{fig:map_rho_orient2}
\end{figure}


In this work we focus on \er ions at the so-called substitution site 1 of the \yso matrix, whose optical transition at 1536.38 nm is well separated from the other crystallographically inequivalent site 2 \cite{bottger_optical_2006_SM}, allowing one to probe only one site. Each \er site has two orientations A and B (sometimes called magnetic sub-sites) with different $g$-tensors connected by a C$_2$-rotation around the $b$-axis of the crystal. The \er orientations are said magnetically equivalent when the magnetic field is applied in the $D_1-D_2$ plane, so they show the same electronic Zeeman splittings, same frequency shift on their optical transitions and therefore are probed simultaneously. We focus on this case for simplicity but also to maximize the total absorption. The magnetic moments of the two orientations are nonetheless oriented differently, leading to very different couplings with their respective surrounding \y ions. For the specific orientation of the external magnetic field chosen here ($225^{\circ}$ from $D_1$), each of the \er orientations is coupled in the exact similar way to one \y ion.

In the main text, we show the map of the maximum branching contrast $\rho_{\rm max}$ for the external magnetic $\vec{B}$ oriented at 225$^{\circ}$ from $D_1$ in the $D_1-D_2$ plane, considering the \er ions of site 1, orientation A. Here, Fig.~\ref{fig:map_rho_orient2} shows the map of $\rho_{\rm max}$ for the same magnetic field configuration but considering the \er ions of site 1, orientation B. Comparing the two maps of $\rho_{\rm max}$, one can notice the similarity of the peaks distribution, in fact related by the transformation  $\theta \rightarrow \pi - \theta$ (C$_2$-rotation around the $b$-axis). The same \y ions are positioned on the map, and now the \y ion with coordinates $(1.01,5.11,1.64)$ \AA~(pinned in red) is located in a region where the branching contrast is potentially large. This \y ion is at the same distance ($r=5.4572$ \AA) from the \er ions than the \y ion identified in the main text (orientation A). Therefore, the interaction between these two \er -\y pairs is the same, and the branching contrast is maximum for the same external magnetic field strength, leading to the exact same transition frequencies $\Delta_g$ and $\Delta_e$ and probabilities. This situation is exceptional because these two particular \y ions are the only ones being linked by a C$_2$-rotation around the $b$-axis among the twenty closest \y ions. In consequence, in the experimental part where we perform photon echo measurements, we optically probe the two \er orientations that undergo the same coupling, each with one \y spin. Two similar echo signals add up to give a total signal that can be interpreted as coming from a single class of \er ions interacting with one \y spin.


\section{Photon echo measurements}

The crystal is 10 ppm \eryso, grown by Scientific Materials Corporation, and with dimensions of $8\times 4\times 3$~mm$^3$. A liquid helium cryostat with variable temperature insert was used to cool down the crystal to $1.8$~K. The magnetic field within the $(D_1,D_2)$ plane of the crystal and at $225^{\circ}$ from $D_1$ is generated via a superconducting magnet. The light is propagating along the 8~mm dimension, parallel to the $b$-axis, and we adjust the polarization along $D_2$ to maximize the absorption \cite{bottger_spectroscopy_2006_SM}.

We use an extended cavity diode 
that is amplified by an EDFA (Manlight), and temporally shaped by an acousto-optic modulator driven at 80 MHz to create two gaussian monochromatic pulses with duration of 110 ns (rms duration), separated by a varying time $t_{12}$. At $t=2t_{12}$, the optical coherences rephase and an echo is emitted. The echo is filtered out from the strong excitation pulses using a gated acousto-optic modulator and detected by an avalanche photodetector (Thorlabs APD110C).


\begin{thebibliography}{10}

\bibitem{childress2006coherent}
L.~Childress, M.~G. Dutt, J.~Taylor, A.~Zibrov, F.~Jelezko, J.~Wrachtrup,
  P.~Hemmer, and M.~Lukin, Science \textbf{314}, 281 (2006).

\bibitem{maurer_room-temperature_2012}
P.~C. Maurer, G.~Kucsko, C.~Latta, L.~Jiang, N.~Y. Yao, S.~D. Bennett,
  F.~Pastawski, D.~Hunger, N.~Chisholm, M.~Markham, D.~J. Twitchen, J.~I.
  Cirac, and M.~D. Lukin, Science \textbf{336}, 1283 (2012).

\bibitem{yang2016high}
S.~Yang, Y.~Wang, D.~B. Rao, T.~H. Tran, A.~S. Momenzadeh, M.~Markham,
  D.~Twitchen, P.~Wang, W.~Yang, R.~St{\"o}hr \emph{et~al.}, Nature Photonics \textbf{10}, 507 (2016).

\bibitem{SiV1}
B.~Pingault, J.~N. Becker, C.~H.~H. Schulte, C.~Arend, C.~Hepp, T.~Godde, A.~I.
  Tartakovskii, M.~Markham, C.~Becher, and M.~Atat\"ure, Phys. Rev. Lett.
  \textbf{113}, 263601 (2014).

\bibitem{SiV2}
L.~J. Rogers, K.~D. Jahnke, M.~H. Metsch, A.~Sipahigil, J.~M. Binder,
  T.~Teraji, H.~Sumiya, J.~Isoya, M.~D. Lukin, P.~Hemmer, and F.~Jelezko, Phys.
  Rev. Lett. \textbf{113}, 263602 (2014).

\bibitem{QD}
M.~Kroutvar, Y.~Ducommun, D.~Heiss, M.~Bichler, D.~Schuh, G.~Abstreiter, and
  J.~J. Finley, Nature \textbf{432}, 81 (2004).

\bibitem{steger_quantum_2012}
M.~Steger, K.~Saeedi, M.~L.~W. Thewalt, J.~J.~L. Morton, H.~Riemann, N.~V.
  Abrosimov, P.~Becker, and H.-J. Pohl, Science \textbf{336}, 1280 (2012).

\bibitem{morley2013quantum}
G.~W. Morley, P.~Lueders, M.~H. Mohammady, S.~J. Balian, G.~Aeppli, C.~W. Kay,
  W.~M. Witzel, G.~Jeschke, and T.~S. Monteiro, Nature materials \textbf{12},
  103 (2013).

\bibitem{Morsee1700930}
K.~J. Morse, R.~J.~S. Abraham, A.~DeAbreu, C.~Bowness, T.~S. Richards,
  H.~Riemann, N.~V. Abrosimov, P.~Becker, H.-J. Pohl, M.~L.~W. Thewalt, and
  S.~Simmons, Science Advances \textbf{3}, e1700930 (2017).

\bibitem{probst}
S.~Probst, H.~Rotzinger, A.~V. Ustinov, and P.~A. Bushev, Phys. Rev. B
  \textbf{92}, 014421 (2015).

\bibitem{bienfait2016reaching}
A.~Bienfait, J.~Pla, Y.~Kubo, M.~Stern, X.~Zhou, C.~Lo, C.~Weis, T.~Schenkel,
  M.~Thewalt, D.~Vion \emph{et~al.}, Nature nanotechnology \textbf{11}, 253
  (2016).

\bibitem{bottger_effects_2009}
T.~B\"ottger, C.~W. Thiel, R.~L. Cone, and Y.~Sun, Physical Review B
  \textbf{79}, 115104 (2009).

\bibitem{dibos_isolating_2017}
A.~Dibos, M.~Raha, C.~Phenicie, and J.~Thompson, arXiv:1711.10368 [physics,
  physics:quant-ph]  (2017). ArXiv: 1711.10368.

\bibitem{Mims_SHF1}
W.~B. Mims, Phys. Rev. B \textbf{5}, 2409 (1972).

\bibitem{Mims_SHF2}
W.~B. Mims, Phys. Rev. B \textbf{6}, 3543 (1972).

\bibitem{mims1965pulsed}
W.~Mims, Proceedings of the Royal Society of London. Series A, Mathematical and
  Physical Sciences \textbf{283}, 452 (1965).

\bibitem{guillot-noel_direct_2007}
O.~Guillot-No\"el, H.~Vezin, P.~Goldner, F.~Beaudoux, J.~Vincent, J.~Lejay, and
  I.~Lorger\'e, Physical Review B \textbf{76}, 180408 (2007).

\bibitem{Ahlefeldt20101594}
R.~Ahlefeldt, W.~Hutchison, and M.~Sellars, Journal of Luminescence
  \textbf{130}, 1594  (2010). Special Issue based on the Proceedings of the
  Tenth International Meeting on Hole Burning, Single Molecule, and Related
  Spectroscopies: Science and Applications (HBSM 2009) - Issue dedicated to
  Ivan Lorgere and Oliver Guillot-Noel.

\bibitem{tiranov_temporal_2016}
A.~Tiranov, P.~C. Strassmann, J.~Lavoie, N.~Brunner, M.~Huber, V.~B. Verma,
  S.~W. Nam, R.~P. Mirin, A.~E. Lita, F.~Marsili, M.~Afzelius, F.~Bussi\`eres,
  and N.~Gisin, Physical Review Letters \textbf{117}, 240506 (2016).

\bibitem{zhong_interfacing_2017}
T.~Zhong, J.~M. Kindem, J.~Rochman, and A.~Faraon, Nature Communications
  \textbf{8}, ncomms14107 (2017).

\bibitem{bottger_optical_2006}
T.~B\"ottger, C.~W. Thiel, Y.~Sun, and R.~L. Cone, Physical Review B
  \textbf{73}, 075101 (2006).

\bibitem{guillot-noel_hyperfine_2006}
O.~Guillot-No\"el, P.~Goldner, Y.~L. Du, E.~Baldit, P.~Monnier, and
  K.~Bencheikh, Physical Review B \textbf{74}, 214409 (2006).

\bibitem{sun_magnetic_2008}
Y.~Sun, T.~B\"ottger, C.~W. Thiel, and R.~L. Cone, Physical Review B
  \textbf{77}, 085124 (2008).
  
  \bibitem{SM_rho}
  See Supplemental Material for the derivation of the branching ratio expression.

\bibitem{SM1}
See Supplemental Material for more
details about the magnetic subsites and the equivalent map for orientation B, which includes \cite{YSOstructure,sun_magnetic_2008,bottger_optical_2006}.

\bibitem{YSOstructure}
B.~A. {Maksimov}, Y.~A. {Kharitonov}, V.~V. {Ilyukhin}, and N.~V. {Belov},
  Soviet Physics Doklady \textbf{13}, 1188 (1969).

\bibitem{Maksimov1970}
B.~A. Maksimov, V.~V. Ilyukhin, Y.~A. Kharitonov, and N.~V. Belov,
  Kristallografiya \textbf{15}, 926 (1970).

\bibitem{SM2}
See Supplemental Material for more details about the experimental setup, which includes \cite{bottger_spectroscopy_2006}.
  
  \bibitem{bottger_spectroscopy_2006}
T.~B\"ottger, Y.~Sun, C.~W. Thiel, and R.~L. Cone, Physical Review B
  \textbf{74}, 075107 (2006).

\bibitem{mitsunaga_cw_1990}
M.~Mitsunaga, Physical Review A \textbf{42}, 1617 (1990).

\bibitem{PhysRevB.7.1834_ruby}
L.~Q. Lambert, Phys. Rev. B \textbf{7}, 1834 (1973).

\bibitem{van1990optically}
E.~Van~Oort and M.~Glasbeek, Chemical Physics \textbf{143}, 131 (1990).

\bibitem{PhysRevB.82.201201}
P.~L. Stanwix, L.~M. Pham, J.~R. Maze, D.~Le~Sage, T.~K. Yeung, P.~Cappellaro,
  P.~R. Hemmer, A.~Yacoby, M.~D. Lukin, and R.~L. Walsworth, Phys. Rev. B
  \textbf{82}, 201201 (2010).

\bibitem{PhysRevB.78.094303}
J.~R. Maze, J.~M. Taylor, and M.~D. Lukin, Phys. Rev. B \textbf{78}, 094303
  (2008).

\bibitem{probst_anisotropic_2013}
S.~Probst, H.~Rotzinger, S.~W\"unsch, P.~Jung, M.~Jerger, M.~Siegel, A.~V.
  Ustinov, and P.~A. Bushev, Physical Review Letters \textbf{110}, 157001
  (2013).

\bibitem{PhysRevLett.114.247603}
A.~L. Falk, P.~V. Klimov, V.~Iv\'ady, K.~Sz\'asz, D.~J. Christle, W.~F. Koehl,
  A.~Gali, and D.~D. Awschalom, Phys. Rev. Lett. \textbf{114}, 247603 (2015).

\bibitem{scheuer2016optically}
J.~Scheuer, I.~Schwartz, Q.~Chen, D.~Schulze-S\"unninghausen, P.~Carl,
  P.~H\"ofer, A.~Retzker, H.~Sumiya, J.~Isoya, B.~Luy \emph{et~al.}, New
  Journal of Physics \textbf{18}, 013040 (2016).

\bibitem{tyryshkin2012electron}
A.~M. Tyryshkin, S.~Tojo, J.~J. Morton, H.~Riemann, N.~V. Abrosimov, P.~Becker,
  H.-J. Pohl, T.~Schenkel, M.~L. Thewalt, K.~M. Itoh \emph{et~al.}, Nature
  materials \textbf{11}, 143 (2012).

\bibitem{obrien_interfacing_2014}
C.~O'Brien, N.~Lauk, S.~Blum, G.~Morigi, and M.~Fleischhauer, Physical Review
  Letters \textbf{113}, 063603 (2014).

\bibitem{fernandez-gonzalvo_coherent_2015}
X.~Fernandez-Gonzalvo, Y.-H. Chen, C.~Yin, S.~Rogge, and J.~J. Longdell,
  Physical Review A \textbf{92}, 062313 (2015).

\bibitem{PhysRevLett.113.203601}
L.~A. Williamson, Y.-H. Chen, and J.~J. Longdell, Phys. Rev. Lett.
  \textbf{113}, 203601 (2014).

\bibitem{kukharchyk2017optical}
N.~Kukharchyk, D.~Sholokhov, S.~L. Korableva, A.~A. Kalachev, and P.~A. Bushev, New Journal of Physics \textbf{20}, 023044 (2018).

\bibitem{ladd_quantum_2010}
T.~D. Ladd, F.~Jelezko, R.~Laflamme, Y.~Nakamura, C.~Monroe, and J.~L.
  O’Brien, Nature \textbf{464}, 45 (2010).

\bibitem{hensen2015loophole}
B.~Hensen, H.~Bernien, A.~E. Dr{\'e}au, A.~Reiserer, N.~Kalb, M.~S. Blok,
  J.~Ruitenberg, R.~F. Vermeulen, R.~N. Schouten, C.~Abell{\'a}n \emph{et~al.},
  Nature \textbf{526}, 682 (2015).

\end{thebibliography}

\begin{thebibliography}{1}

\bibitem{YSOstructure_SM}
B.~A. {Maksimov}, Y.~A. {Kharitonov}, V.~V. {Ilyukhin}, and N.~V. {Belov},
  Soviet Physics Doklady \textbf{13}, 1188 (1969).

\bibitem{sun_magnetic_2008_SM}
Y.~Sun, T.~B\"ottger, C.~W. Thiel, and R.~L. Cone, Physical Review B
  \textbf{77} (2008).

\bibitem{bottger_optical_2006_SM}
T.~B\"ottger, C.~W. Thiel, Y.~Sun, and R.~L. Cone, Physical Review B
  \textbf{73}, 075101 (2006).

\bibitem{bottger_spectroscopy_2006_SM}
T.~B\"ottger, Y.~Sun, C.~W. Thiel, and R.~L. Cone, Physical Review B
  \textbf{74}, 075107 (2006).

\end{thebibliography}
\end{document}